\documentclass[pdflatex,sn-basic]{sn-jnl}% Basic Springer Nature Reference Style/Chemistry Reference Style

\usepackage{graphicx}%
\usepackage{multirow}%
\usepackage{amsmath,amssymb,amsfonts}%
\usepackage{amsthm}%
\usepackage{mathrsfs}%
\usepackage[title]{appendix}%
\usepackage{xcolor}%
\usepackage{textcomp}%
\usepackage{manyfoot}%
\usepackage{booktabs}%
\usepackage{algorithm}%
\usepackage{algorithmicx}%
\usepackage{algpseudocode}%
\usepackage{listings}%
\usepackage{caption} % Preamble'a ekleyin
\usepackage{geometry}
\usepackage{siunitx}
\usepackage{subcaption}

\theoremstyle{thmstyleone}%
\theoremstyle{thmstyletwo}%
\theoremstyle{thmstylethree}%

\usepackage{multicol}
\usepackage{lipsum} % örnek metin için
\usepackage{tabularx}
\restylefloat{table}
\begin{document}

\title{\textbf{Predictive Modeling of High-Altitude Clear Air Turbulence in the United States: A Machine Learning Approach}}


\author{\textbf{Kadir Gökdeniz}$^1$(corresponding author) ORCID: 0009-0004-0901-8642 kadirqokdeniz@hotmail.com· \textbf{İrem Ülkü}$^2$ ORCID: 0000-0003-4998-607X}


%\affil*[1]{\orgdiv{Department}, \orgname{Organization}, \orgaddress{\street{Street}, \city{City}, \postcode{100190}, \state{State}, \country{Country}}}

%\affil[2]{\orgdiv{Department}, \orgname{Organization}, \orgaddress{\street{Street}, \city{City}, \postcode{10587}, \state{State}, \country{Country}}}

%\affil[3]{\orgdiv{Department}, \orgname{Organization}, \orgaddress{\street{Street}, \city{City}, \postcode{610101}, \state{State}, \country{Country}}}


\abstract{High-altitude Clear Air Turbulence (CAT) poses significant risks to aviation safety due to its unpredictability and challenges in detection. This study leverages machine learning models to improve CAT prediction within U.S. airspace at 200–350 hPa pressure levels, utilizing Pilot Reports (PIREPs), ERA5 reanalysis data, and aircraft aerodynamic parameters from the BADA database. Gradient boosting algorithms, particularly XGBoost, achieved the highest performance with an AUC of 0.904, demonstrating superior capability in capturing non-linear atmospheric dynamics. Key findings highlight the dominance of geographic coordinates (17.5\% feature importance) and turbulence indices like TI3 in prediction, emphasizing the role of regional topography and upper-tropospheric instability. The integration of aerodynamic features such as drag force and wing loading improved the detection of moderate-to-severe perceived turbulence intensity (POD improved from 0.845 to 0.866), providing additional value to traditional aircraft-independent methods. Seasonal analysis revealed winter months as peak periods for CAT incidents, correlating with jet stream activity. While results align with global studies, limitations include geographic scope and aircraft-type diversity. This research underscores the potential of machine learning for operational CAT forecasting, with recommendations for future work focusing on global data integration and real-time telemetry to address climate-driven turbulence trends.}




\maketitle
\begin{multicols}{2}
\section*{Introduction}
The safety and efficiency of aviation operations are directly influenced by meteorological conditions, which pose significant operational risks. Factors such as turbulence, low visibility, low-level wind shear, precipitation, icing, low cloud ceilings, volcanic ash, and sandstorms can disrupt flight schedules and pose risks to both human and aircraft safety.

Turbulence stands as one of the most critical factors affecting aviation safety, presenting significant risks to both passengers and aircraft. A study by Gultepe et al. reveals that 71\% of aviation accidents caused by atmospheric conditions are attributed to turbulence (Gultepe et al. 2019). Moreover, turbulence is estimated to impose an annual economic cost of \$100 to \$200 million on the United States (Paul D. et al. 2017). However, turbulence is one of the most complex challenges in fluid dynamics due to its inherently chaotic and unpredictable nature. While the Navier-Stokes equations provide a solid theoretical foundation, solving these equations requires substantial computational power, limiting their practical applications (Wang Rui et al. 2020). Consequently, numerical weather predictions and various ensemble methods have been developed, but the need for innovative approaches remains.

Turbulence can be classified into several subtypes, each with unique dynamics. For instance, mountain wave turbulence occurs when wind crosses mountain ridges, creating a wavy airflow that poses significant risks to low-altitude flights (Sharman R. 2016). Convective turbulence, on the other hand, typically arises on hot days with intense vertical air movement or near thunderstorm clouds (Nilsson et al. 2001).

Clear Air Turbulence (CAT) is a phenomenon that occurs unexpectedly in clear skies and is challenging to observe directly by pilots or detect via radar systems (Venkatesh et al. 2013). This makes CAT one of the most unpredictable types of turbulence, granting it a special focus in aviation research. Notably, recent research suggests a rising trend in clear air turbulence across regions including East Asia, the Eastern Pacific, and the Northwestern Pacific. (Lee et al. 2022). Furthermore, a strong correlation has been highlighted between global warming and clear air turbulence, emphasizing the role of climate-induced atmospheric instabilities in this increase (Foudad 2024). Another study by Meneguz et al. shows that airflow over mountain ranges in the upper troposphere can elevate the likelihood of encountering CAT and orographic turbulence to over 80\% (Meneguz E. et al. 2016).

Historically, numerical weather prediction (NWP) models have been the primary approach for analyzing clear air turbulence (CAT). (Kim et al. 2021, Kristoffer 2013). In recent years, numerical weather prediction has been integrated with machine learning algorithms
\end{multicols}

\begin{table*}[!h]
\centering
\begin{tabularx}{\linewidth}{c*{8}{>{\raggedright\arraybackslash}X}} 
    & Reference & Region & Methods & Time Interval & Dataset & Time Resolution & Types of Turbulence & Results \\
    \midrule
    & Muñoz-Esparza et al. (2020) & USA & Random Forest, GBRT & June 2018- September 2019 & \footnotemark[1] and \footnotemark[7] & 6 hours & CAT, MWT & AUC: 0.770 - 0.906 \\
    \hline
    & Hon et al. (2020) & Asia-Pacific & XGBoost & April 2018 - March 2019 & \footnotemark[1] and \footnotemark[2] & 1 hour & MOG CAT &  AUC score: 0.84 \\
    \hline
    & Shao et al. (2024) & China & XGBoost, SVM, LR, RF, DT & Jan 2020 - Dec 2022 & \footnotemark[1],\footnotemark[2] and ERA5  & 1 hour & MOG CAT ,CIT & AUC scores: 0.75 - 0.96  \\
    \hline
    & Nerushev et al. (2022) & Northern Hemisphere & Correlation-extreme algorithms & 2007-2018 & \footnotemark[3],\footnotemark[4] and \footnotemark[5] & 15 minutes (image interval) & CAT, MWT & R$^{2}$ = 0.68 - 0.74 \\
    \hline
    & Sharman et al. (2016) & United States & Statistical Mapping of Turbulence & --- & \footnotemark[2], GFS, \footnotemark[6] and ERA5  & 6 hours & CAT, MWT & AUC: 0.77 - 0.91 \\
    \hline
    & Pearson et al. (2017) & US National Airspace & Integrate \footnotemark[1], \footnotemark[2], \footnotemark[8], \footnotemark[9], \footnotemark[10] & Jul-Sep 2015, Jan-Mar 2016 & \footnotemark[2], \footnotemark[8], EDR, WRF-RAP & 15 minutes & Convective, CAT, MWT & AUC: 0.75 - 0.88 \\
    \hline
    & Storer et al. (2020) & Airspace of Europe and the UK & \footnotemark[11] & Sep 2016 - Aug 2017 & Boeing 747 and 777 aircraft & 24, 27, 30, and 33 hours & ST, MTW, and CIT & AUC: 0.84 - 0.88  \\
    \hline
    & de Mello (2024) & Brazil & Random Forest & Mar 2018 - Mar 2021 & GFS, ERA-5, VRTG & --- & CAT (3 different intensities) & POD:0.74- 0.78, FAR: 0.26 -0.31 \\
\end{tabularx}
\caption{Literature Review Summary}
\label{tab:Literature Review}
\end{table*}

\begin{multicols}{2}

\footnotetext[1]{In situ observation}
\footnotetext[2]{PIREP}
\footnotetext[3]{SEVIRI Radiometer Data}
\footnotetext[4]{NCEP/NCAR Reanalysis Data}
\footnotetext[5]{NOAA Arctic Sea Ice Data}
\footnotetext[6]{WRF-RAP}
\footnotetext[7]{HRRR Predictions}
\footnotetext[8]{METAR}
\footnotetext[9]{NTDA}
\footnotetext[10]{DCIT}
\footnotetext[11]{Multi-diagnostic multi-model ensemble}

\noindent for CAT forecasting, demonstrating their effectiveness (Hon et al. 2020, Muñoz-Esparza et al. 2020). Studies in the literature exhibit significant variability in terms of problem definitions, geographical application areas, study periods, data sources, resolution levels, turbulence intensities, and findings. Table 1 categorizes and summarizes some studies on CAT analysis based on various criteria.

Muñoz-Esparza et al. (2020) conducted a study on the weaknesses of the GTG method and proposed a machine learning approach based on Random Forest instead. The study applied EDR transformations and utilized various turbulence indicators derived from HRRR forecast data. For 6-hour forecasts of turbulence types such as CAT and MWT, the results achieved an AUC score of 0.906.

A study by Hon et al. (2020) focuses on using the Multi Index Consensus (MIC) method for turbulence prediction in the Asia-Pacific region (Hon et al. 2020). The research processes over 16,000 pilot report (PIREP) data and 19 different turbulence diagnostic features using the XGBoost algorithm, demonstrating an improvement in prediction performance ranging from 3\% to 17\%. This study highlights the benefits of applying machine learning techniques to local-scale turbulence forecasting.

The study conducted by Shao et al. (2024) applied five different machine learning methods to predict Moderate or Greater (MOG) CAT turbulence over China (Shao et al. 2024). By utilizing in situ observations, PIREPs, and the ERA5 dataset, the research analyzed the temporal and spatial characteristics of turbulence through 15 distinct turbulence diagnostics.

This study conducted by Nerushev et al.2022 analyzes turbulence areas in the upper troposphere using satellite data from 2007-2018, investigating the relationship between turbulence and jet streams and climatic parameters (Nerushev et al.2022). The study identifies turbulence areas through water vapor inhomogeneities and correlation-extremal algorithms and uses regression models to explain the temporal variability of strong turbulence areas.

In their 2016 study, Sharman and Pearson developed a statistical mapping method for turbulence forecasting (Sharman et al.2016). This method converts various turbulence indicators derived from NWP model outputs into the ICAO-standard EDR. The study evaluated the model's accuracy by utilizing ROC curves for different turbulence categories. The results demonstrated that ensemble averaging (GTG) achieved higher AUC scores and outperformed individual indicators.

The study by Pearson and Sharman in 2017 is a continuation of their 2016 work (Pearson et al. 2017). In the initial research, the performance of the GTG community model was enhanced by integrating data sources such as DCIT, NTDA, PIREP, and METAR, leading to the creation of a new model called GTGN. It was highlighted that GTGN performs better in capturing convective turbulence, especially during the summer months.

Storer et al. 2020 investigated multi-diagnostic, multi-model ensemble forecasting for aviation turbulence, finding that combining multiple diagnostics improved skill over single-model ensembles (Storer et al. 2020). MOGREPS-G outperformed EPS51, and EPS12 slightly improved results. Smaller ensemble spreads within multi-diagnostic ensembles yielded better performance, with further study needed on optimal ensemble size.

The study conducted by de Mello 2024 applies machine learning algorithms using GFS, ERA-5, and VRTG data for Clear Air Turbulence (CAT) prediction, covering part of the Brazilian border. The Random Forest algorithm demonstrated the best performance in CAT prediction, achieving POD of 0.74 and FAR of 0.31 with GFS data, and POD of 0.78 and FAR of 0.26 with ERA-5 data.

In this study, the PIREP and ERA5 datasets were used to analyze data with a 1-hour time resolution from 2022 to 2024 within the United States. The study focused on CAT classification and examined upper altitude data at 200, 225, 250, 300, and 350 hPA levels. The performance of XGBoost, RandomForest, LightGBM, AdaBoost, and Logistic Regression models was compared.

This study differs from the existing literature primarily by examining how incorporating aircraft type and aerodynamic features from pilot reports into the model could potentially enhance classification performance. This approach, unlike traditional aircraft-independent classification methods, addresses the problem from an aircraft-specific perspective and models the turbulence intensity as perceived by the pilots.
A comprehensive evaluation of high-altitude CAT samples obtained from various data sources presents a systematic approach to overcome limitations arising from the subjective nature of pilot reports.

\section*{Data}

\subsection*{PIREP}

The IEM (Iowa Environmental Mesonet) database consists of PIREP (Pilot Reports) data, which contains information reported by pilots during their flights about atmospheric conditions they encounter. In this study, data obtained from the IEM database was utilized. For proper matching with meteorological data, information on latitude, longitude, altitude, and time of each turbulence observation is required. The IEM database provides all this information and also contains details about the types and severity of turbulence reported by pilots.
In this study, the data was filtered according to specific criteria:

\begin{enumerate}
    \item Only high-altitude data at pressure levels of 200, 225, 250, 300, and 350 hPA was selected.
    \item Data spanning from 2022 through 2024 was used.
    \item Geographically, the US airspace between 50°N and 24°N latitude, and -125°E and -67°E longitude was examined.
    \item Among turbulence types, only Clear Air Turbulence (CAT) related data was selected.
\end{enumerate}

This filtering allowed the study to focus specifically on Clear Air Turbulence cases occurring at high altitudes within US airspace.

Between 2022 and 2024, a total of 19,213 Clear Air Turbulence (CAT) cases were recorded. These CAT events were classified into five different categories based on their intensity: LGT, LGT-MOD, MOD, SEV-MOD, SEV.

Creating a balanced dataset is crucial for the effective application of machine learning models.Therefore, to balance the dataset, 19,213 NEG (negative - no turbulence) cases were also selected to match the 19,213 positive CAT cases, resulting in a balanced dataset
\end{multicols}

\begin{figure}[!b]
    \centering
    \includegraphics[width=\columnwidth]{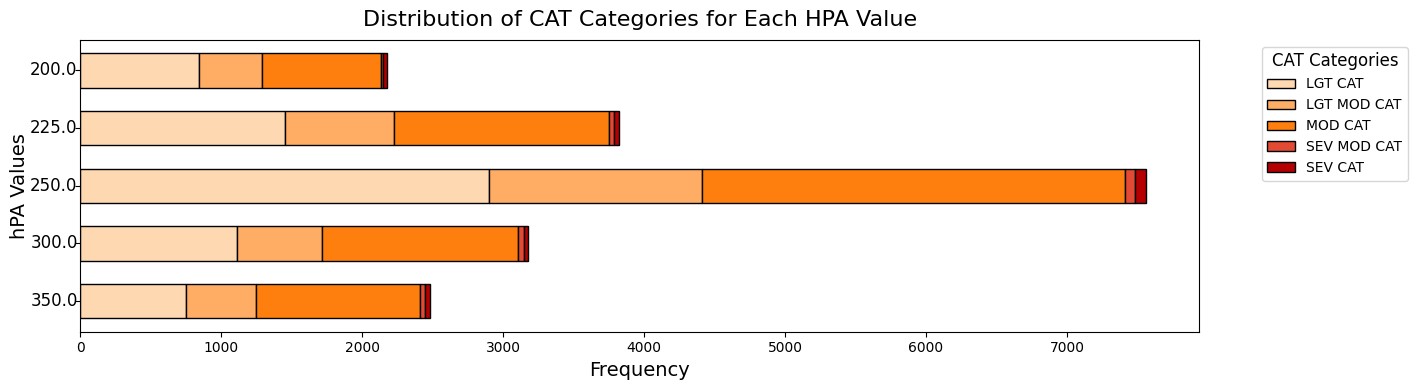}
    \caption{Distribution of CAT Categories for Each HPA Value in US Airspace between 2022 and 2024.}
    \label{fig:hpa_distribution} % Etiketi değiştirdim
\end{figure}

\begin{figure}[!t]
    \vspace*{-\topskip}
    \centering
    \includegraphics[width=\columnwidth]{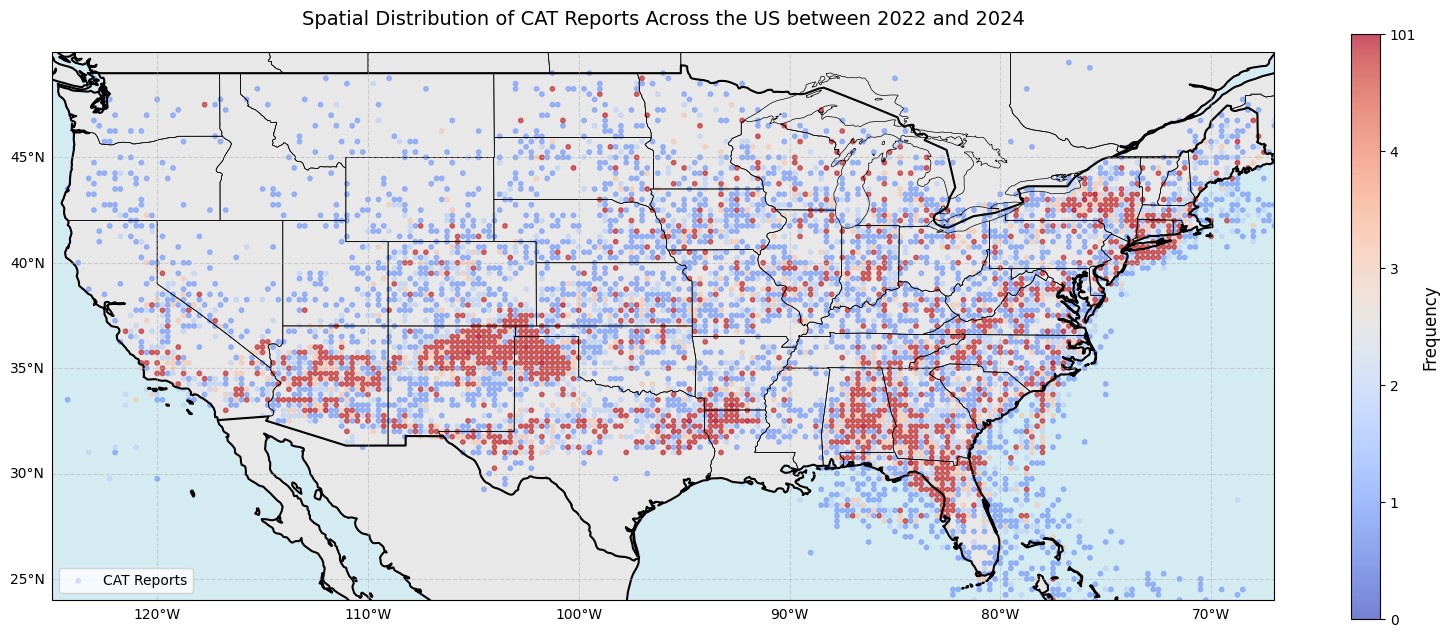}
    \caption{Spatial Distribution of CAT Reports Across the US between 2022 and 2024.}
    \label{fig:spatial_distribution} % Etiketi değiştirdim
\end{figure}

\begin{multicols}{2}
\noindent consisting of a total of 38,426 cases.

Figure 1, shows the distribution of CAT events by severity categories recorded at different pressure levels between 2022 and 2024. The highest frequency of CAT events is observed at the 250 hPA level, where all severity categories are represented. A particularly notable point is that MOD CAT events are significantly higher at the 250 hPA level.

Figure 2, illustrates the spatial distribution of CAT reports across US airspace between 2022 and 2024. The visualization reveals distinct regional clustering of CAT events.Notably, high-frequency CAT reports are concentrated in the Southwest region (Arizona, New Mexico, and surrounding areas),the South (intersection of Texas, Oklahoma, and Arkansas), along the Eastern seaboard (from the Carolinas to Florida), and in the Northeast corridor (particularly around New York). This distribution likely correlates with the mountainous terrain features, jet stream pathways, and air traffic density in these regions.

Figure 3, illustrates the seasonal distribution of turbulence categories from 2022 to 2024. The graph clearly demonstrates a seasonal cycle in turbulence reports. Winter and spring months show a marked increase in MOD CAT reports, with the 2023-2024 winter season reaching the highest number of MOD CAT reports at approximately 1100 incidents. LGT CAT reports follow a similar seasonal pattern, tending to increase during winter and spring months. SEV CAT and SEV MOD CAT categories remain relatively low throughout the year but exhibit slight increases during winter seasons.

Summer months show a notable decrease in all turbulence categories, suggesting that summer conditions are less conducive to CAT formation.This seasonal pattern can be attributed to jet stream activity and stronger atmospheric frontal systems during winter 

\end{multicols}

\begin{figure}[!b]
    \centering
    \includegraphics[width=\columnwidth]{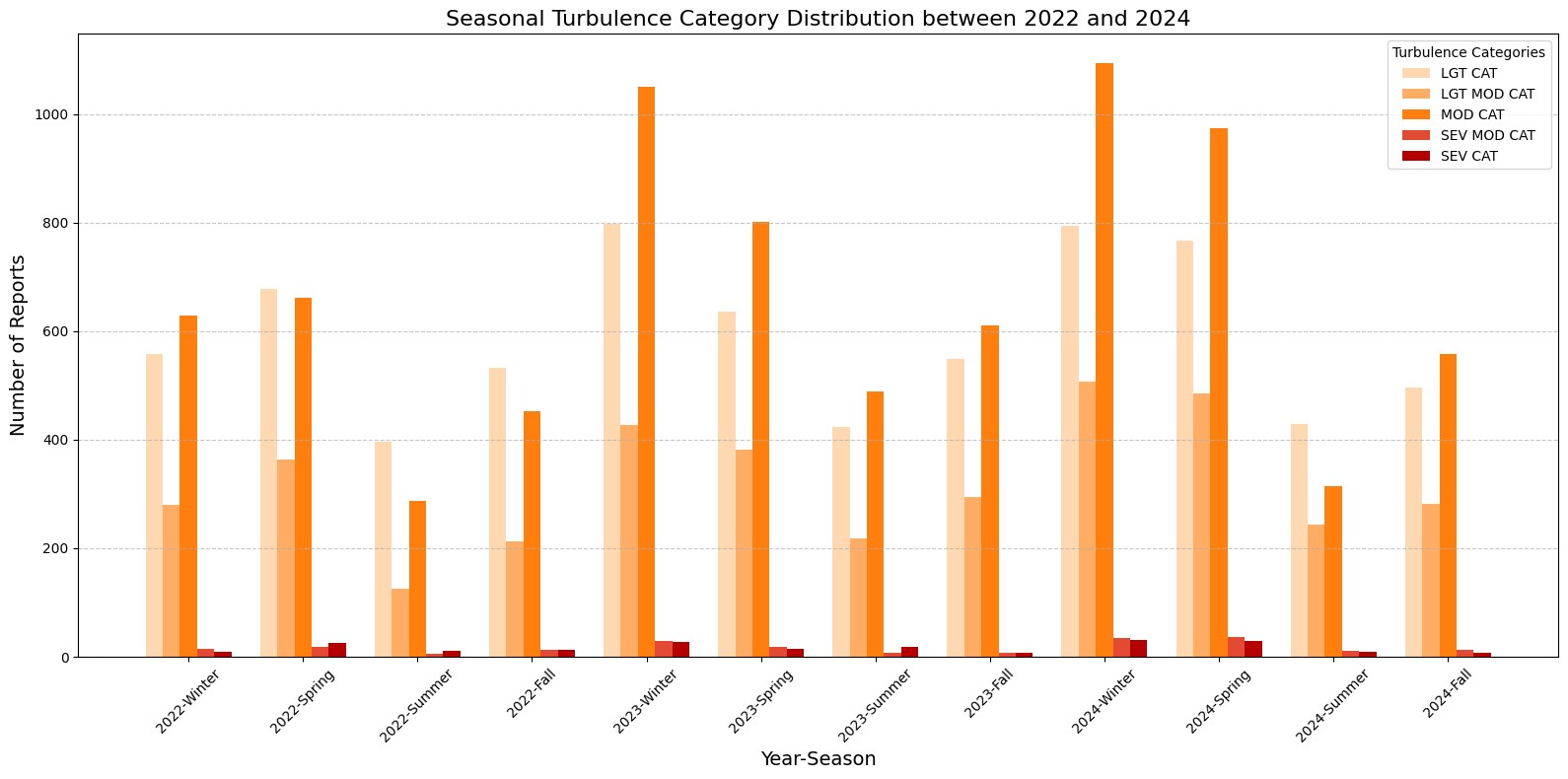}
    \caption{Seasonal Turbulence Category Distribution Across the US between 2022 and 2024.}
    \label{fig:seasonal_distribution} % Etiketi değiştirdim
\end{figure}

\begin{figure}[!t]
    \centering
    \includegraphics[width=\columnwidth]{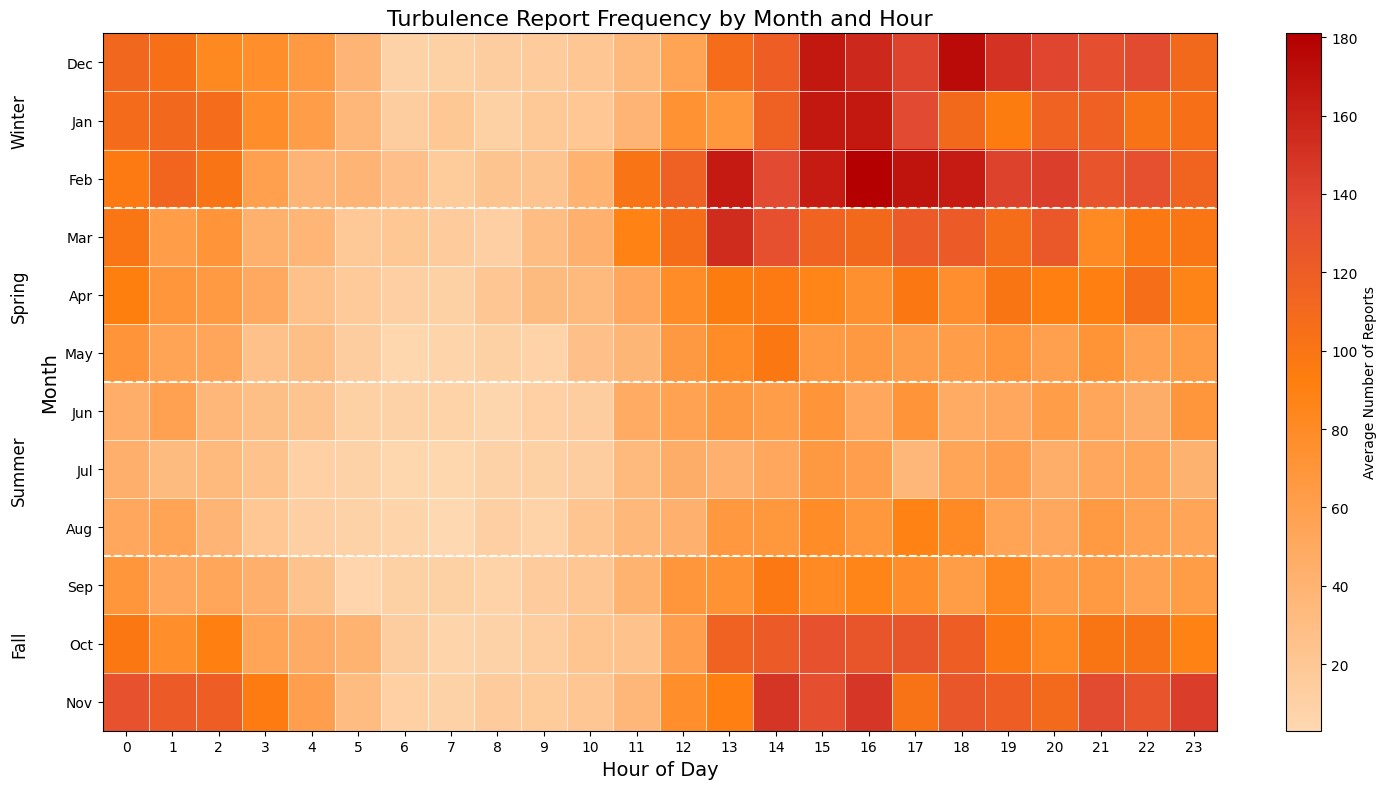}
    \caption{Turbulence Report Frequency by Month and Hour}
    \label{fig:monthly_hourly_distribution} % Etiketi değiştirdim
\end{figure}

\begin{multicols}{2}

\noindent months.

Figure 4, illustrates the frequency of turbulence reports throughout the year, broken down by month and hour of day.
Winter months (December, January, February) show a significant increase in turbulence reports, particularly during evening hours (15:00-21:00). During these periods, the number of reports reaches 140-180.Across all months, early morning hours and late afternoon/evening hours show increased turbulence reports. The period between 14:00-20:00 is consistently the most intense. Summer months (June, July, August) and midday hours (6:00-11:00) represent the periods with the fewest turbulence reports. Spring (March-May) and fall (September-November) months show a transition between winter and summer patterns. Turbulence reports begin to increase in November and start decreasing from March onward.

Based on this data, flight planning recommendations could include exercising additional caution during evening hours in winter months, while summer months and midday hours could be considered as periods with lower expected turbulence.

\subsection*{ERA 5 Data}

The ERA 5 pressure level dataset is frequently preferred in turbulence analysis studies (Shao et al. 2024, Sharman et al. 2016, de Mello 2024).
The ERA5 reanalysis dataset has a spatial resolution of 0.25 degrees latitude by 0.25 degrees longitude and a temporal resolution of 1 hour. This research examines pressure levels of 200, 225, 250, 300, and 350 hPa, which correspond to the typical high-altitude flight levels used by aircraft. The study covers data from 2022 to 2024 and focuses on the region between 50°N and 24°N latitude, and -125°E and -67°E longitude.

The pressure levels database in ECMWF includes important variables related to turbulence, such as geopotential, relative humidity, temperature, potential vorticity, and wind components (U, W, V).

Due to the complex nature of turbulence, rather than attempting to achieve a complete representation of turbulence, there is a need for indicators that signify the presence of turbulence. The turbulence indicators used in this study are comprehensively summarized in Table 2. These indicators, based on various atmospheric parameters, allow us to evaluate the potential for turbulence.The selected parameters can be fundamentally categorized into two main groups: dynamic wind components and thermal effects.

Parameters representing thermal effects include components such as Theta (potential temperature) and TR (temperature gradient/Richardson number). These parameters measure temperature changes in the atmosphere, modeling the impact of thermal instabilities on turbulence formation. Regions where temperature gradients change abruptly are particularly critical for clear air turbulence (CAT) formation.

Dynamic wind components include parameters such as wind shear, Dutton index, and DVT, which characterize both spatial and temporal variations of wind. These parameters measure the complex dynamic behaviors of air masses in the atmosphere, reflecting the energy potential necessary for turbulence formation.

Exploratory data analysis revealed that turbulence diagnostic parameters in the ERA5 dataset exhibit a non-Gaussian distribution. Figure 5 illustrates the distribution of turbulence diagnostics across severity levels using box plots. For comparative visualization purposes only, we normalized the values in this figure, while maintaining non-normalized raw data throughout the actual study.

The most significant finding in the analysis results is the different distribution patterns of turbulence parameters across severity categories. Parameters such as $n2$,$wing\ loading\ ref$, and $t$ exhibit high normalized values across all turbulence intensity levels, demonstrating their fundamental position in turbulence characterization regardless of severity level. In contrast, parameters including $richardson$, $vorsq$, and $ti3$ show notably low values, with the Richardson number's consistently low values confirming its known inverse relationship with atmospheric instability and turbulence potential. The increasing values of wind-related parameters such as $wind\ shear$ from MOD CAT to MOD SEV CAT indicate their potential importance in distinguishing between severity levels, while revealing the complex, non-linear structure in turbulence manifestation.

The non-normalized raw data used in our study preserves the actual physical values of the parameters and their natural relationships with each other. This approach enables the creation of better representations of the physical foundations of turbulence formation mechanisms.

\subsection*{BADA}

BADA (Base of Aircraft Data) is a comprehensive aircraft performance model developed and maintained by EUROCONTROL. This database contains detailed operational characteristics for various aircraft types, including performance parameters, aerodynamic coefficients, and engine thrust properties. 

This study examines the performance benefits of incorporating aircraft type and aerodynamic features from pilot reports into turbulence classification models. Pilot reports typically include aircraft type information, which enables direct matching with corresponding aerodynamic properties in the BADA database. This aircraft-specific approach enables the model to account for how different aircraft types experience and respond to the same atmospheric conditions, potentially providing predictions of perceived turbulence.

\section*{Aircraft-Specific Aerodynamic Parameters}

This study analyzes the influence of aircraft-specific aerodynamic parameters on turbulence response. We utilize BADA (Base of Aircraft Data) parameters to calculate key aerodynamic metrics, establishing relationships between aircraft type and turbulence intensity perception.

\end{multicols}

\begin{table}[!t]
\centering
\renewcommand{\arraystretch}{1.5}
\resizebox{\textwidth}{!}{%
\begin{tabular}{ccccp{5cm}}
\hline
Diagnostic & Description & Unit & Algorithm & Reference \\
\hline
$\text{Ri}$ & Negative Richardson number & — & 
$\text{Ri} = -\frac{N^2}{\text{VWS}^2}$ & 
Oard (1974) \\
\hline
TI3$^a$ & \begin{tabular}{c}Variant 3 of Elrod's \\ turbulence index\end{tabular} & $\text{s}^{-1}$ &$\text{TI3} = \text{TI1} + \text{DVT}$ & Lee et. al. (2022) \\
\hline
PVGRAD$^b$ & Potential vorticity gradient & $\text{PVU km}^{-1}$
  & $\text{PV}_{\text{grad}} = \sqrt{150\left(\nabla_h \text{PV}\right)^2 + \left(\frac{\partial \text{PV}}{\partial z}\right)^2}$ &  Homeyer et. al. (2021) \\
\hline
VORSQ & Relative vorticity squared & $\text{s}^{-2}$ & 
$\text{VORSQ} = \left| \frac{\partial v}{\partial x} - \frac{\partial u}{\partial y} \right|^2$ & 
Sharman et al. (2006) \\
\hline
$\text{DVT}^c$ & Divergence Tendency & $\text{s}^{-1}$  & $\text{DVT} = C \left| \left(\frac{\partial u}{\partial x} + \frac{\partial v}{\partial y}\right)_{t2} - \left(\frac{\partial u}{\partial x} + \frac{\partial v}{\partial y}\right)_{t1} \right|$ & Lee et. al. (2022) \\
\hline
Dutton$^d$ & Dutton's empirical index & — & 
$\text{Dutton} = 1.25|\text{HWS}| + 0.25\text{VWS}^2 + 10.5$ & 
Dutton (1980) \\
\hline
$|\text{DIV}|$ & \begin{tabular}{c}Magnitude of \\ horizontal divergence\end{tabular} & $\text{s}^{-1}$ & 
$|\text{DIV}| = \left| \frac{\partial u}{\partial x} + \frac{\partial v}{\partial y} \right|$ & 
Sharman et al. (2006) \\
\hline
TR & \begin{tabular}{c}Horizontal temperature \\ gradient/Richardson number\end{tabular} & $\text{K m}^{-1}$ & 
$\text{TR} = \frac{\left| \frac{\partial t}{\partial x} + \frac{\partial t}{\partial y} \right|}{Ri}$ & 
Kim et al. (2018) \\
\hline
$\theta$ & Potential temperature & K & 
$\theta = T \left( \frac{p_0}{p} \right)^{R/c_p}$ & 
Holton et. al. (2012) \\
\hline
Wspeed & Wind speed & $\text{m s}^{-1}$ & 
$\text{Wspd} = (u^2 + v^2)^{1/2}$ & 
Endlich (1964) \\
\hline
TI2$^e$ & \begin{tabular}{c}Variant 2 of Elrod's \\ turbulence index\end{tabular} & $\text{s}^{-2}$ & 
$\text{TI2} = \text{VWS} \times (\text{DEF-DIV})$ & 
Elrod and Knapp (1992) \\
\hline
$\text{WINDS/RI}^f$ & Wind Speed$^2$/Richardson Number & $\text{m}^2 \text{s}^{-2}$ & $\text{WINDS/RI} = \text{Wspeed}^2/\text{Ri}$ & Holton et. al. (2012) \\
\hline
VWS & \begin{tabular}{c}Vertical shear of \\ horizontal wind\end{tabular} & $\text{s}^{-1}$ & 
$\text{VWS} = \left[ \left( \frac{\partial u}{\partial z} \right)^2 + \left( \frac{\partial v}{\partial z} \right)^2 \right]^{1/2}$ & 
Endlich (1964) \\
\hline
N² & \begin{tabular}{c}Brunt-Väisälä \\ frequency squared\end{tabular} & $\text{s}^{-2}$ & 
$\text{N}^2 = \frac{g}{\theta} \frac{\partial \theta}{\partial z}$ & 
Sharman et al. (2006) \\
\hline
\end{tabular}
}
\caption{Atmospheric turbulence diagnostic parameters and their formulations.}
\label{tab:turbulence_parameters}

\raggedright
\footnotesize
In this table, several parameters require additional clarification. The variable TI1 refers to Elrod's original turbulence index, calculated as the product of vertical wind shear (VWS) and flow deformation (DEF). Potential vorticity (PV) is defined by the equation $-g(\frac{\partial \theta}{\partial p})(\frac{\partial v}{\partial x} - \frac{\partial u}{\partial y} + f)$, where $f$ represents the Coriolis parameter. For the divergence tendency (DVT) calculation, a constant scaling factor of $C = 0.01$ is applied. Horizontal wind shear (HWS) is expressed as $\frac{1}{V^2}(v\frac{\partial u}{\partial x} - u\frac{\partial u}{\partial y} + v\frac{\partial v}{\partial x} - u\frac{\partial v}{\partial y})$, while flow deformation (DEF) equals $\sqrt{\text{DSH}^2 + \text{DST}^2}$, combining stretching deformation (DSH = $\frac{\partial u}{\partial x} - \frac{\partial v}{\partial y}$) and shearing deformation (DST = $\frac{\partial v}{\partial x} + \frac{\partial u}{\partial y}$).
\end{table}

\begin{figure}[!b]
    \centering
    \includegraphics[width=\columnwidth]{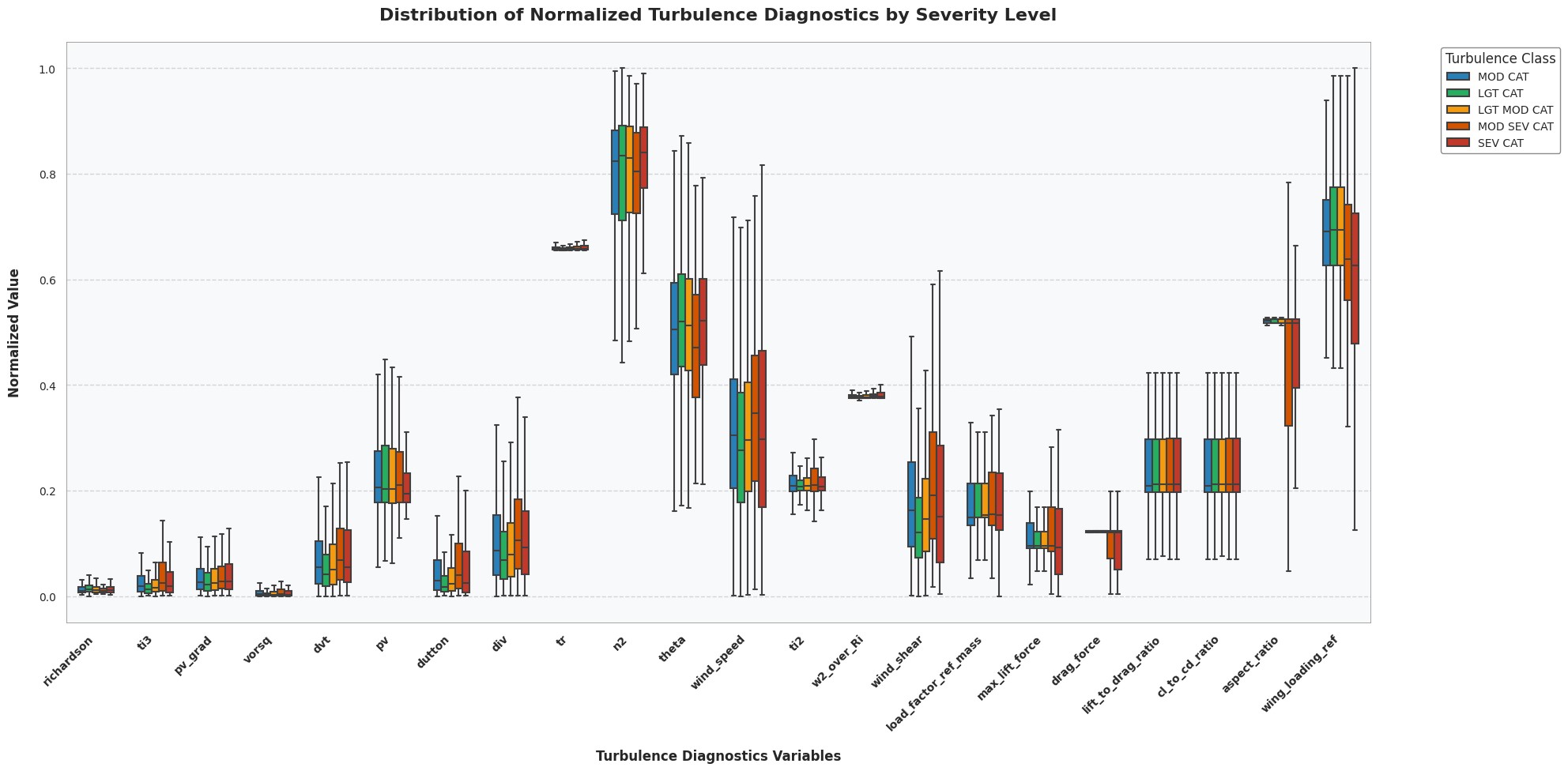}
    \caption{Turbulence Report Frequency by Month and Hour}
    \label{fig:normalized_distribution} % Etiketi değiştirdim
\end{figure}

\begin{multicols}{2}

\subsection*{Mathematical Formulation}

The aircraft-specific parameters were calculated using the following equations:

\begin{align}
L_{max} &= \frac{1}{2} \rho V^2 S (C_{L,bo} \cdot B_c) \tag{Maximum lift force} \\
n_{ref} &= \frac{L_{max}}{m_{ref} \cdot g} \tag{Load factor} \\
F_d &= \frac{1}{2} \rho V^2 S C_d \tag{Drag force} \\
\frac{L}{D} &= \frac{L_{max}}{F_d} \tag{Lift-to-drag ratio} \\
\frac{C_L}{C_d} &= \frac{C_{L,bo} \cdot B_c}{C_d} \tag{Lift-to-drag coefficient ratio} \\
AR &= \frac{b^2}{S} \tag{Aspect ratio} \\
\frac{W}{S} &= \frac{m_{ref} \cdot g}{S} \tag{Wing loading}
\end{align}

Where:
\begin{itemize}
    \item $\rho$: Air density [kg/m$^3$]
    \item $V$: True airspeed [m/s]
    \item $S$: Wing area [m$^2$]
    \item $C_{L,bo}$: Basic lift coefficient
    \item $B_c$: Buffet coefficient
    \item $m_{ref}$: Reference mass [kg]
    \item $g$: Gravitational acceleration [9.81 m/s$^2$]
    \item $C_d$: Drag coefficient
    \item $b$: Wingspan [m]
\end{itemize}

\section*{Methods}

\subsection*{Turbulence Diagnostic Parameters and Data Preparation}

For the input of our machine learning model, various turbulence diagnostic parameters with proven significance in aviation literature have been selected. Comprehensive descriptions of these parameters are presented in detail in Table 2. Data obtained from Pilot Reports (PIREPs) provides latitude, longitude, altitude, and time information, and turbulence diagnostic parameters in ERA5 are labeled with these. In the modeling phase, instances where Clear Air Turbulence (CAT) was present were labeled as "1" and instances where it was absent were labeled as "0" using a binary classification approach.

One of the distinctive aspects of our study is the inclusion of aircraft aerodynamic properties in the model, in addition to traditional turbulence diagnostic parameters. This integration was accomplished using data provided from EUROCONTROL's BADA (Base of Aircraft Data) database. The aircraft type information provided by PIREPs was matched with corresponding aerodynamic properties in the BADA database, enabling a more comprehensive turbulence prediction model.

\subsection*{Data Categorization and Analysis Scenarios} 

In our research, we categorized our dataset in various ways to more deeply understand the effect of aerodynamic properties on turbulence detection. The data was divided into two categories based on turbulence intensity: "light to light moderate" and "moderate to severe." Additionally, cases where aerodynamic properties were included and cases where they were excluded were analyzed separately.When all these categorizations and general performance analyses are combined, a total of six different scenarios were examined in detail in our study.

In addition to RandomForest and XGBoost algorithms, which are predominantly used for CAT diagnosis in the literature, we also implemented AdaBoost and LightGBM algorithms. Furthermore, Logistic Regression was examined as a potential alternative to decision tree algorithms.The entire dataset was randomly split into 70\% training and 30\% testing data for model training and evaluation, and all models were analyzed

\end{multicols}

\begin{table}[!h]
    \centering
    \caption{Metrics used for evaluating model performance.}
    \label{tab:performance_metrics}
    \begin{tabular}{lll}
        \hline
        \textbf{Metric} & \textbf{Formula} & \textbf{Description} \\
        \hline
        POD (Probability of Detection) & $\frac{TP}{TP+FN}$ & Rate of correctly detected turbulence events. Higher \\
        & & values indicate better detection of hazardous areas, \\
        & & enhancing flight safety. \\
        \hline
        FAR (False Alarm Ratio) & $\frac{FP}{TP+FP}$ & Proportion of false turbulence predictions. Lower \\
        & & values reduce unnecessary route changes and \\
        & & improve operational efficiency. \\
        \hline
        CSI (Critical Success Index) & $\frac{TP}{TP+FP+FN}$ & Overall success in predicting turbulence events, \\
        & & considering both false alarms and missed events. \\
        & & Balances safety and operational efficiency. \\
        \hline
        TSS (True Skill Statistic) & $\frac{TP}{TP+FN} - \frac{FP}{FP+TN}$ & Balanced measure of prediction ability for both \\
        & & turbulent and non-turbulent conditions. Ranges \\
        & & from -1 to 1, with 1 being perfect prediction. \\
        & & Effective for imbalanced datasets. \\
        \hline
    \end{tabular}
\end{table}

\begin{multicols}{2}
\noindent using these data partitions.

\subsection*{Model Evaluation}

For a single classifier model, the outputs can be categorized into four types: True Positive (TP), False Positive (FP), True Negative (TN), and False Negative (FN). These outputs are used as calculation metrics for ROC (Receiver Operating Characteristic) curves and AUC (Area Under Curve) to evaluate model performance. This approach is frequently preferred in the literature for evaluating CAT prediction models (Cai et al. 2023; Hu et al. 2022; Kim et al. 2018; Lee et al. 2020; Sharman et al. 2006).

In addition, POD (Probability of Detection), FAR (False Alarm Ratio), CSI (Critical Success Index), and TSS (True Skill Statistic) metrics, which are commonly used in the literature for CAT prediction, were also employed in the evaluation (Sharman et al. 2016, Pearson et al. 2017, Mu˜noz-
Esparza et al. 2020 ). The formulations for these metrics are presented in detail in Table 3.

POD indicates the rate at which turbulence events that actually occur are correctly detected by the model. A high POD value enhances flight safety by ensuring pilots are informed of potentially hazardous areas.

FAR represents the proportion of events predicted as turbulence
by the model  but which do not actually occur. A low FAR value improves operational efficiency by preventing unnecessary route changes and fuel consumption.

CSI comprehensively evaluates the model's success in correctly predicting turbulence events. It takes into account both false alarms and missed turbulence events, thus reflecting the balance between safety and efficiency in aviation operations.

TSS assesses the model's ability to correctly predict both turbulent and non-turbulent air conditions in a balanced manner. This metric is particularly important in imbalanced datasets like turbulence prediction, as turbulence events occur infrequently.

The AUC score of 0.5 represents the lower performance limit producing random results, while a score of 1 represents the ideal classifier. In this context, the performance of the models developed for high-altitude CAT diagnosis has been comprehensively evaluated using these metrics.

\section*{Results}
In this study, the performance of XGBoost, LightGBM,
\end{multicols}

\begin{table}[!b]
\centering
\caption{Performance Comparison of Machine Learning Models Without and With Aerodynamic Features}
\label{tab:model_performance}
\scriptsize
\setlength{\tabcolsep}{3pt}
\begin{tabular}{ll|ccccc|ccccc}
\toprule
\multirow{2}{*}{\textbf{Turbulence}} & \multirow{2}{*}{\textbf{Model}} & \multicolumn{5}{c|}{\textbf{Without Aerodynamic Features}} & \multicolumn{5}{c}{\textbf{With Aerodynamic Features}} \\
\cmidrule(lr){3-7} \cmidrule(lr){8-12}
 &  & \textbf{POD} & \textbf{FAR} & \textbf{CSI} & \textbf{TSS} & \textbf{AUC} & \textbf{POD} & \textbf{FAR} & \textbf{CSI} & \textbf{TSS} & \textbf{AUC} \\
\midrule
\multirow{6}{*}{\begin{tabular}[c]{@{}l@{}}All\\Categories\end{tabular}} 
 & Random Forest & 0.790 & 0.175 & 0.677 & 0.623 & 0.887 & 0.785 & 0.170 & 0.676 & 0.624 & 0.889 \\
 & XGBoost & 0.796 & \textbf{0.154} & \textbf{0.695} & \textbf{0.652} & \textbf{0.901} & 0.809 & \textbf{0.158} & \textbf{0.703} & \textbf{0.657} & \textbf{0.904} \\
 & LightGBM & \textbf{0.800} & 0.160 & 0.694 & 0.648 & 0.900 & \textbf{0.815} & 0.166 & 0.701 & 0.652 & 0.902 \\
 & AdaBoost & 0.790 & 0.161 & 0.692 & 0.641 & 0.893 & 0.770 & 0.161 & 0.682 & 0.633 & 0.893 \\
 & Logistic Regression & 0.633 & 0.247 & 0.524 & 0.425 & 0.767 & 0.624 & 0.273 & 0.505 & 0.389 & 0.734 \\
\midrule
\multirow{6}{*}{MOD-SEV} 
 & Random Forest & 0.828 & 0.141 & 0.728 & 0.691 & 0.919 & 0.849 & 0.155 & 0.735 & 0.693 & 0.919 \\
 & XGBoost & 0.845 & \textbf{0.139} & 0.743 & 0.708 & \textbf{0.928} & \textbf{0.866} & 0.150 & \textbf{0.753} & \textbf{0.716} & \textbf{0.928} \\
 & LightGBM & \textbf{0.878} & 0.161 & \textbf{0.751} & \textbf{0.709} & \textbf{0.928} & 0.857 & 0.154 & 0.741 & 0.701 & 0.923 \\
 & AdaBoost & 0.848 & 0.149 & 0.742 & 0.702 & 0.921 & 0.839 & \textbf{0.140} & 0.741 & 0.695 & 0.923 \\
 & Logistic Regression & 0.682 & 0.205 & 0.580 & 0.506 & 0.816 & 0.651 & 0.211 & 0.554 & 0.477 & 0.786 \\
\midrule
\multirow{6}{*}{LGT-LGT MOD} 
 & Random Forest & 0.790 & 0.204 & 0.656 & 0.587 & 0.870 & 0.758 & \textbf{0.188} & 0.645 & 0.582 & 0.872 \\
 & XGBoost & \textbf{0.802} & 0.196 & \textbf{0.670} & \textbf{0.606} & \textbf{0.881} & \textbf{0.803} & 0.197 & \textbf{0.671} & \textbf{0.606} & \textbf{0.884} \\
 & LightGBM & 0.788 & \textbf{0.190} & 0.665 & 0.603 & 0.880 & 0.790 & 0.189 & 0.667 & \textbf{0.606} & 0.879 \\
 & AdaBoost & 0.779 & 0.200 & 0.651 & 0.579 & 0.868 & 0.789 & 0.203 & 0.660 & 0.594 & 0.871 \\
 & Logistic Regression & 0.598 & 0.261 & 0.494 & 0.386 & 0.737 & 0.569 & 0.267 & 0.471 & 0.361 & 0.715 \\
\bottomrule
\end{tabular}
\end{table}

\begin{table}[!t]
\centering
\caption{Algorithm performance metrics across different turbulence intensity categories}
\label{tab:turbulence_metrics}
\begin{tabular}{l@{\hspace{7cm}}ccccc}
\hline
\textbf{Turbulence Category} & \textbf{POD} & \textbf{FAR} & \textbf{CSI} & \textbf{TSS} & \textbf{AUC} \\
\hline
Neg & & & & & \\
$\sigma_{\text{XGBoost}}$ & 0.001 & 0.001 & 0.003 & 0.003 & 0.001 \\
\hline
LGT & & & & & \\
$\sigma_{\text{XGBoost}}$ & 0.005 & 0.002 & 0.001 & 0.004 & 0.002 \\
\hline
LGT-MOD & & & & & \\
$\sigma_{\text{XGBoost}}$ & 0.004 & 0.003 & 0.001 & 0.003 & 0.011 \\
\hline
MOD & & & & & \\
$\sigma_{\text{XGBoost}}$ & 0.000 & 0.003 & 0.001 & 0.001 & 0.002 \\
\hline
SEV-MOD & & & & & \\
$\sigma_{\text{XGBoost}}$ & 0.013 & 0.020 & 0.000 & 0.023 & 0.002 \\
\hline
SEV & & & & & \\
$\sigma_{\text{XGBoost}}$ & 0.011 & 0.020 & 0.001 & 0.022 & 0.003 \\
\hline
\end{tabular}
\begin{tablenotes}
\small
\item \textbf{Note:} To evaluate model stability and generalization capability, we implemented a k-fold cross-validation analysis for the baseline XGBoost model. Specifically, a fivefold approach was utilized (k = 5), maintaining consistency with our primary data partitioning strategy of 70\% for training and 30\% for testing datasets. This methodology allowed us to systematically assess the variability of statistical performance metrics across different data subsets, providing a robust evaluation of the model's predictive reliability.
\end{tablenotes}
\end{table}

\begin{figure}[!b]
    \centering
    \includegraphics[width=\columnwidth]{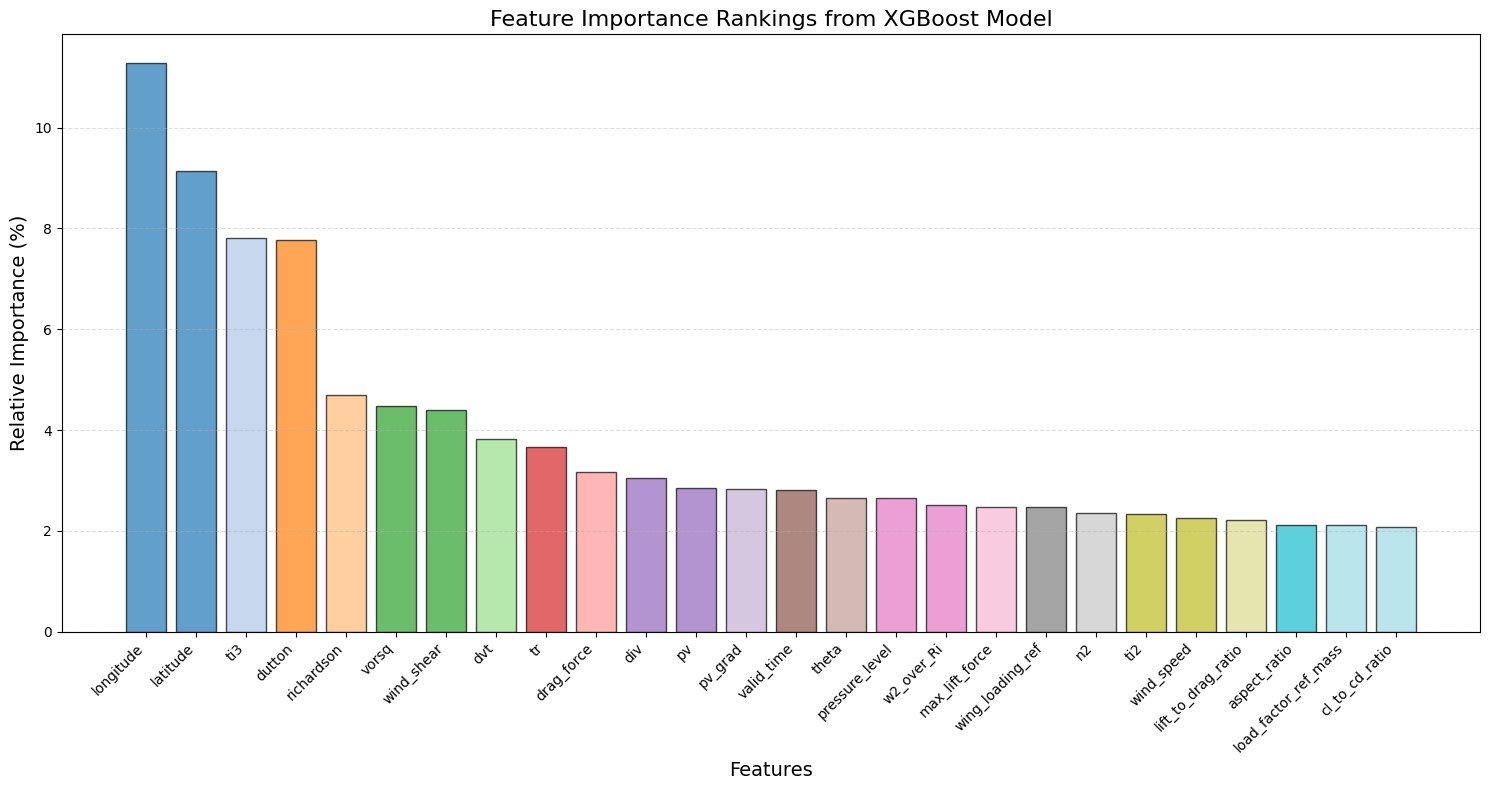}
    \caption{All Categories XGBoost Relative Feature Importance}
    \label{fig:cat_distribution}
\end{figure}

\begin{multicols}{2}

\noindent Random Forest, AdaBoost, and Logistic Regression models for high-altitude clear air turbulence (CAT) prediction was comprehensively examined, taking into account the integration of aerodynamic features. The findings presented in Table 4 clearly demonstrate that gradient boosting-based models (XGBoost and LightGBM) exhibited the highest performance across all categories. XGBoost, which had an AUC value of 0.901 before the addition of aerodynamic features, maintained its leading position with an AUC value of 0.904 after the integration of these features. Particularly in the moderate to severe turbulence (MOD-SEV) category, the increase in the model's POD (Probability of Detection) value from 0.845 to 0.866 provided a significant improvement in the detection of dangerous turbulence events. This improvement contributes to more accurate detection of turbulence events by reducing false negative rates, presenting an important advancement toward enhancing operational safety.

The LightGBM algorithm also demonstrated similarly impressive performance. With the addition of aerodynamic features, the POD value in "All Categories" increased from 0.800 to 0.815, and the TSS (True Skill Statistic) rose from 0.648 to 0.652, exhibiting a balanced prediction capability. These results once again confirm the superiority of gradient boosting methods in modeling the non-linear dynamics of turbulence.

While the integration of aerodynamic features did not provide a consistent reduction in FAR (False Alarm Ratio) values overall, supportive effects on operational efficiency were observed in some models. For example, the AdaBoost algorithm successfully reduced FAR from
\end{multicols}

\begin{figure}[!t]
    \centering
    \includegraphics[width=\columnwidth]{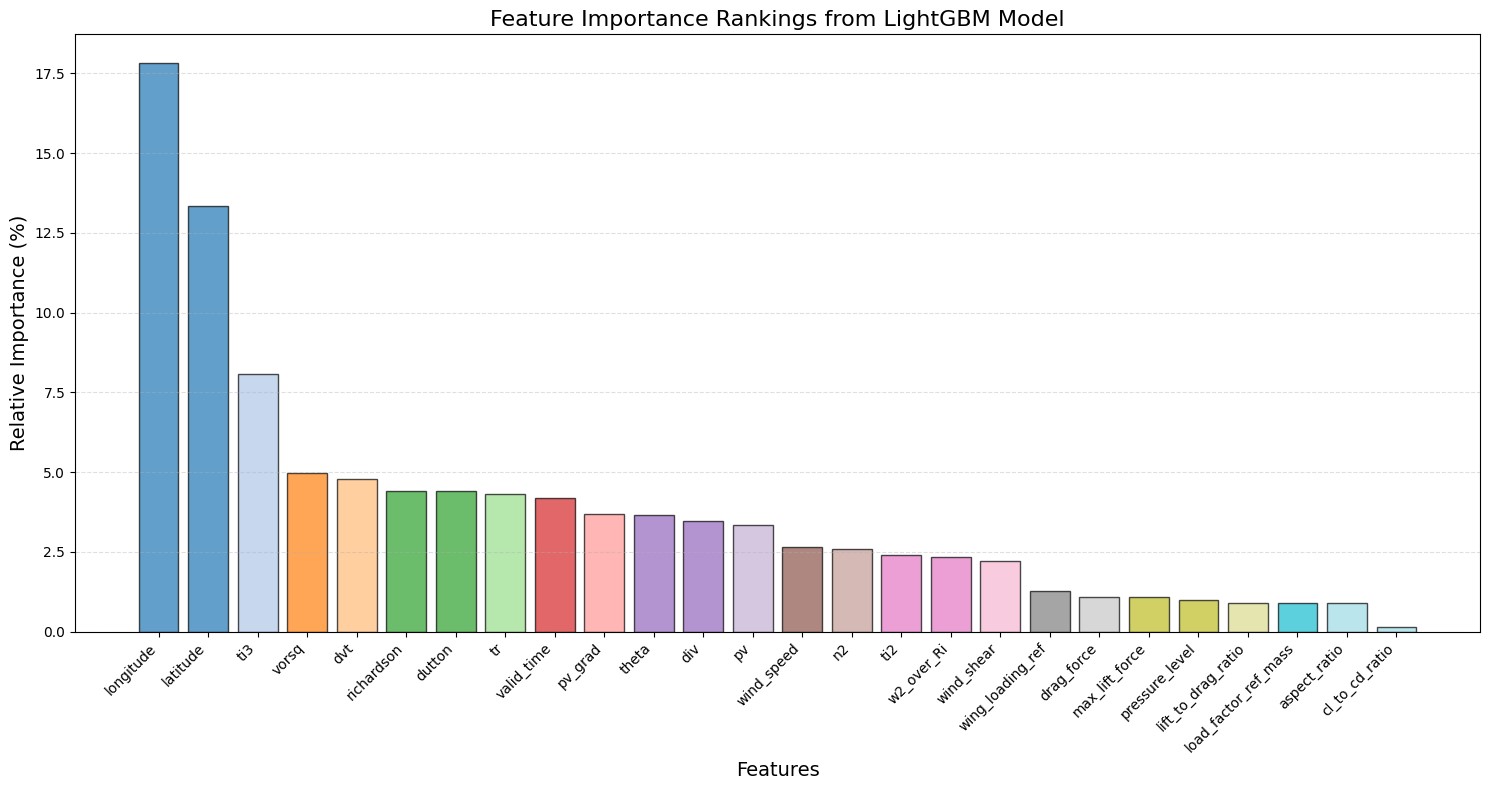}
    \caption{All Categories LightGBM Relative Feature Importance}
    \label{fig:cat_distribution}
\end{figure}

\begin{figure}[!b]
    \centering
    \includegraphics[width=\columnwidth]{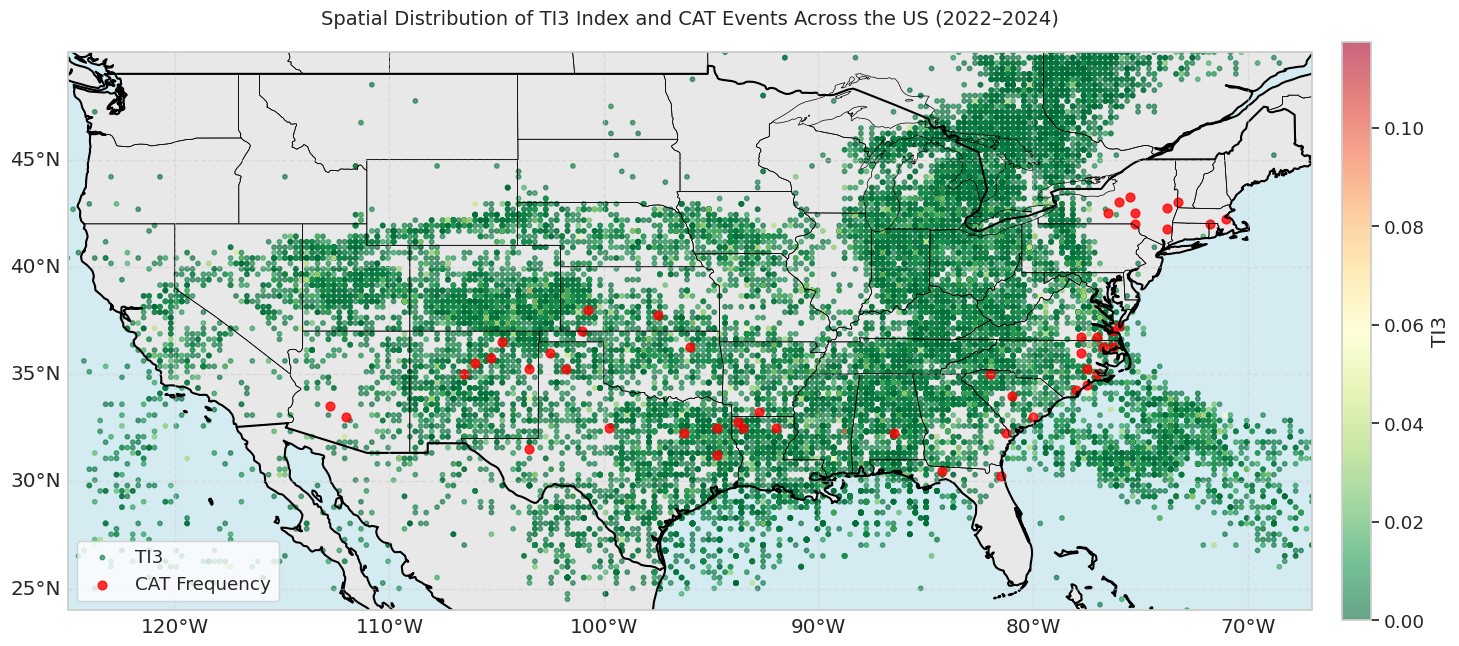}
    \caption{Spatial Distribution of Ti3 Index and CAT Events Across the US between 2022 and 2024}
    \label{fig:cat_distribution}
\end{figure}

\begin{multicols}{2}

\noindent 0.149 to 0.140 in the "MOD-SEV" category,demonstrating success in controlling false alarms. Similarly, the Random Forest model recorded a marginal improvement by reducing FAR from 0.175 to 0.170 in "All Categories." However, the increase of FAR from 0.154 to 0.158 in XGBoost shows that these features may affect the model's specificity in some cases.

The performance of Logistic Regression remained significantly lower compared to other models. The decrease of the AUC value from 0.767 to 0.734 in "All Categories" clearly demonstrates that the complex physical structure of turbulence cannot be captured with linear models. This situation once again highlights the inevitable superiority of ensemble methods in atmospheric events dominated by non-linear relationships.

The standard deviation values presented in Table 5 demonstrate that the performance metrics of the XGBoost model are quite consistent during the cross-validation process ($\sigma_{AUC}$: 0.001--0.003). According to the results in Table 4, the 0.021 increase in POD and 0.003 improvement in AUC observed in the MOD-SEV category with the addition of aerodynamic features are significantly larger than the corresponding standard deviations ($\sigma_{POD}$: 0.013; $\sigma_{AUC}$: 0.002). This supports that these improvements are statistically significant. Changes in FAR values are similar in magnitude to the standard deviations ($\sigma_{FAR}$: 0.020), suggesting that fluctuations in this parameter may be due to random variation. In conclusion, the significance of the increases in the model's key performance indicators emphasizes the effectiveness of aerodynamic feature integration.

Figures 6 and 7 show the feature importance rankings of LightGBM and XGBoost models, respectively. In both models, geographic coordinates (longitude and latitude) stand out as the most determinant features. Longitude has a relative importance of 17.5\% in LightGBM and 11\% in XGBoost. This demonstrates that turbulence events have a strong spatial relationship with topographic structures and regional meteorological dynamics.

While the atmospheric stability indicator \textit{ti3} parameter ranks third in both models (LightGBM: 8.0\%; XGBoost: 7.8\%), the \textit{dutton} index related to vertical wind shear is notable in XGBoost with an importance level of 7.8\%. These findings indicate that atmospheric dynamic factors play a fundamental role in turbulence prediction.

Interestingly, while aerodynamic parameters are present in both models, they generally show relatively modest importance scores below 2.5\%. Among these, \textit{drag\_force} ranks highest (3.2\% in XGBoost), followed by \textit{max\_lift\_force} and \textit{wing\_loading\_ref}. "This suggests that while aerodynamic features contribute to the predictive capabilities of the models, they provide supplementary rather than primary information for perceived turbulence intensity. The consistency between LightGBM and XGBoost importance rankings reinforces the reliability of these findings; however, XGBoost shows a more balanced distribution of importance across features. This suggests that, as shown in Table 4, the XGBoost algorithm can extract some information from aerodynamic features that affect pilots' perception of turbulence.

Figure 8 presents the spatial relationship between the Ti3 index and Clear Air Turbulence events based on pilot reports in US airspace during the 2022-2024 period. Red dots represent regions where CAT reports are concentrated, while green dots represent areas where the Ti3 index is high. The distribution of these two parameters shows a distinct spatial overlap, except for around the New York area; this supports the physical connection between atmospheric dynamics and CAT.

\section*{Conclusion and Discussion}
This study evaluates the performance of machine learning models in predicting high-altitude clear air turbulence (CAT) within the 200–350 hPa pressure levels and U.S. airspace. The AUC value of 0.904 achieved by the gradient boosting algorithm XGBoost is consistent with studies employing similar methodologies (e.g., Muñoz-Esparza et al., 2020: AUC 0.906), supporting its potential integration into operational systems. The dominant role of geographic coordinates (17.5\%) and the Ti3 index clearly reflects CAT's relationship with topography and upper-tropospheric dynamics. This finding aligns with prior research emphasizing the criticality of regional meteorological factors in CAT formation (Muñoz-Esparza et al., 2020).

While the relatively limited contribution of aerodynamic parameters (3.2\%) underscores the continued practicality of traditional aircraft-agnostic approaches, the increase in POD to 0.866 for MOD-SEV categories suggests their complementary role in risk management. The observed seasonal rise in CAT incidents during winter months, linked to jet stream activity and atmospheric instability, complements Shao et al. (2024)’s focus on convective systems.

Key limitations include the dataset's restricted geographic scope and lack of aircraft type diversity. Future research could improve model generalizability through global-scale data integration and real-time telemetry. Additionally, innovative data balancing techniques to augment severe turbulence samples are recommended.

This study demonstrates the efficacy of machine learning models in high-altitude CAT prediction, highlighting the superiority of gradient boosting algorithms (XGBoost and LightGBM) in capturing non-linear atmospheric relationships. The limited yet strategic contribution of aerodynamic parameters offers quantitative advantages for early detection of severe turbulence in operational systems. By providing a data-driven approach to manage climate change-induced CAT increases, this research indicates pathways for further refinement through global data integration and algorithmic advancements.

\textbf{\textit{Funding}} This research received no specifc grant from any funding
agency in the public, commercial, or not-for-proft sectors.

\textbf{\textit{Acknowledgments}} We would like to thank the Presidency of Defence Industries of the Republic of Turkey and the SAYZEK-ATP program for the support and guidance provided within the scope of this study. The EUROCONTROL BADA dataset used in the study was provided by our technical mentor Tolga Baştürk as part of the SAYZEK-ATP program.

\textbf{\textit{Data Availability}} ERA5 reanalysis data are publicly available from the Copernicus Climate Data Store (https://cds.climate.copernicus.eu/). PIREP data are publicly available from the Iowa Environmental Mesonet database (https://mesonet.agron.iastate.edu/request/gis/pireps. php). BADA aircraft performance data were provided through EUROCONTROL under academic research license.

\textbf{\textit{Competing Interests}} The authors declare no competing financial or non-financial interests.

\section*{References}
Gultepe, I., Sharman, R., Williams, P.D. et al. A Review of High Impact Weather for Aviation Meteorology. Pure Appl. Geophys. 176, 1869–1921 (2019). https://doi.org/10.1007/s00024-019-02168-6

Williams, P.D. Increased light, moderate, and severe clear-air turbulence in response to climate change. Adv. Atmos. Sci. 34, 576–586 (2017). https://doi.org/10.1007/s00376-017-6268-2

WANG, Rui, et al. Towards physics-informed deep learning for turbulent flow prediction. Proceedings of the 26th ACM SIGKDD international conference on knowledge discovery \& data mining. 1457-1466 (2020). https://doi.org/10.1145/3394486.3403198

Sharman, R. D., and J. M. Pearson, 2017: Prediction of Energy Dissipation Rates for Aviation Turbulence. Part I: Forecasting Nonconvective Turbulence. J. Appl. Meteor. Climatol., 56, 317–337, https://doi.org/10.1175/JAMC-D-16-0205.1.

Pearson, J. M., and R. D. Sharman, 2017: Prediction of Energy Dissipation Rates for Aviation Turbulence. Part II: Nowcasting Convective and Nonconvective Turbulence. J. Appl. Meteor. Climatol., 56, 339–351, https://doi.org/10.1175/JAMC-D-16-0312.1.

Nilsson, E. D., Rannik, Ü., Kumala, M., Buzorius, G., \& O’dowd, C. D. (2001). Effects of continental boundary layer evolution, convection, turbulence and entrainment, on aerosol formation. Tellus B: Chemical and Physical Meteorology, 53(4), 441–461. https://doi.org/10.3402/tellusb.v53i4.16617

VENKATESH, T.N., MATHEW, J. The problem of clear air turbulence: Changing perspectives in the understanding of the phenomenon. Sadhana 38, 707–722 (2013). https://doi.org/10.1007/s12046-013-0161-1

Lee, J. H., Kim, J.-H., Sharman, R. D., Kim, J., \& Son, S.-W. (2023). Climatology of Clear-Air Turbulence in upper troposphere and lower stratosphere in the Northern Hemisphere using ERA5 reanalysis data. Journal of Geophysical Research: Atmospheres, 128, e2022JD037679. https://doi.org/10.1029/2022JD037679

Foudad, M., Sanchez‐Gomez, E., Jaravel, T., Rochoux, M. C., \& Terray, L. (2024). Past and future trends in clear‐air turbulence over the northern hemisphere. Journal of Geophysical Research: Atmospheres, 129, e2023JD040261. https://doi.org/10.1029/2023JD040261

Meneguz, Elena \& Wells, Oak \& Turp, Debi. (2016). An Automated System to Quantify Aircraft Encounters with Convectively Induced Turbulence over Europe and the North-East Atlantic. Journal of Applied Meteorology and Climatology. 55. 160229122124002. https://doi.org/10.1175/JAMC-D-15-0194.1. 

Muñoz-Esparza, D., Sharman, R. D., \& Deierling, W. (2020). Aviation Turbulence Forecasting at Upper Levels with Machine Learning Techniques Based on Regression Trees. Journal of Applied Meteorology and Climatology, 59(11), 1883-1899. https://doi.org/10.1175/JAMC-D-20-0116.1

Kai Kwong Hon, Cho Wing Ng, Pak Wai Chan,Machine learning based multi-index prediction of aviation turbulence over the Asia-Pacific,Machine Learning with Applications, Volume 2, 2020, 100008, ISSN 2666-8270, https://doi.org/10.1016/j.mlwa.2020.100008.

Shao, J., Z. Zhuang, Z. Yu, K. Lin, K. Wu, Y. Y. Leung, and P. W. Chan, 2024: The Prospective Application of Machine Learning in Turbulence Forecasting over China. J. Appl. Meteor. Climatol., 63, 1273–1285, https://doi.org/10.1175/JAMC-D-23-0236.1.

A F Nerushev et al 2022 IOP Conf. Ser.: Earth Environ. Sci. 1040 012025DOI 10.1088/1755-1315/1040/1/012025

Storer, Luke N., Philip G. Gill, and Paul D. Williams. "Multi‐diagnostic multi‐model ensemble forecasts of aviation turbulence." Meteorological Applications 27.1 (2020): e1885.

de Mello, Ivan Bitar Fiuza, Gutemberg Borges França, and Haroldo Fraga de Campos Velho. "Enhancing Clear Air Turbulence Prediction: A Comparative Analysis of Machine Learning Algorithms Using GFS Forecast and ERA-5 Reanalysis Data." (2024).

Homeyer, C. R., \& Bowman, K. P. (2021). A 22-Year evaluation of convection reaching the stratosphere over the United States. Journal of Geophysical Research: Atmospheres, 126, e2021JD034808. https://doi.org/10.1029/2021JD034808

Dutton, M. J. O., 1980: Probability forecasts of clear-air turbulence based on numerical output. Meteor. Mag., 109, 293–319.

Holton, James R., and Gregory J. Hakim. An Introduction to Dynamic Meteorology. 5th ed. Academic Press, 2013.

Endlich, R. M., 1964: The mesoscale structure of some regions of clear-air turbulence. J. Appl. Meteor., 3, 261–276, https://doi.org/10.1175/1520-0450(1964)003,0261:TMSOSR.2.0.CO;2.

Ellrod, G. P., and D. I. Knapp, 1992: An objective clear-air turbulence forecasting technique: Verification and operational use. Wea. Forecasting, 7, 150–165, https://doi.org/10.1175/1520-0434 (1992) 007,0150:AOCATF.2.0.CO;2.

\end{multicols}

\begin{figure}[!h]
    \centering
    \begin{subfigure}[b]{0.48\textwidth}
        \includegraphics[width=0.90\textwidth]{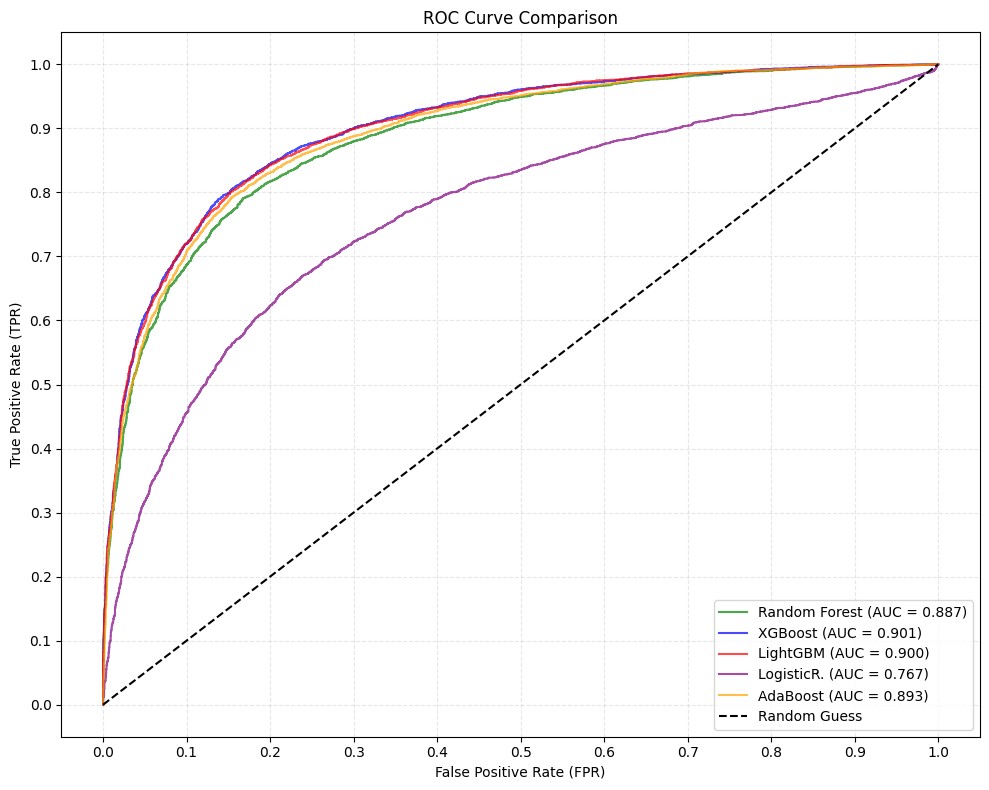}
        \caption{All categories without aerodynamic features}
        \label{fig:roc1}
    \end{subfigure}
    \hfill
    \begin{subfigure}[b]{0.48\textwidth}
        \includegraphics[width=0.90\textwidth]{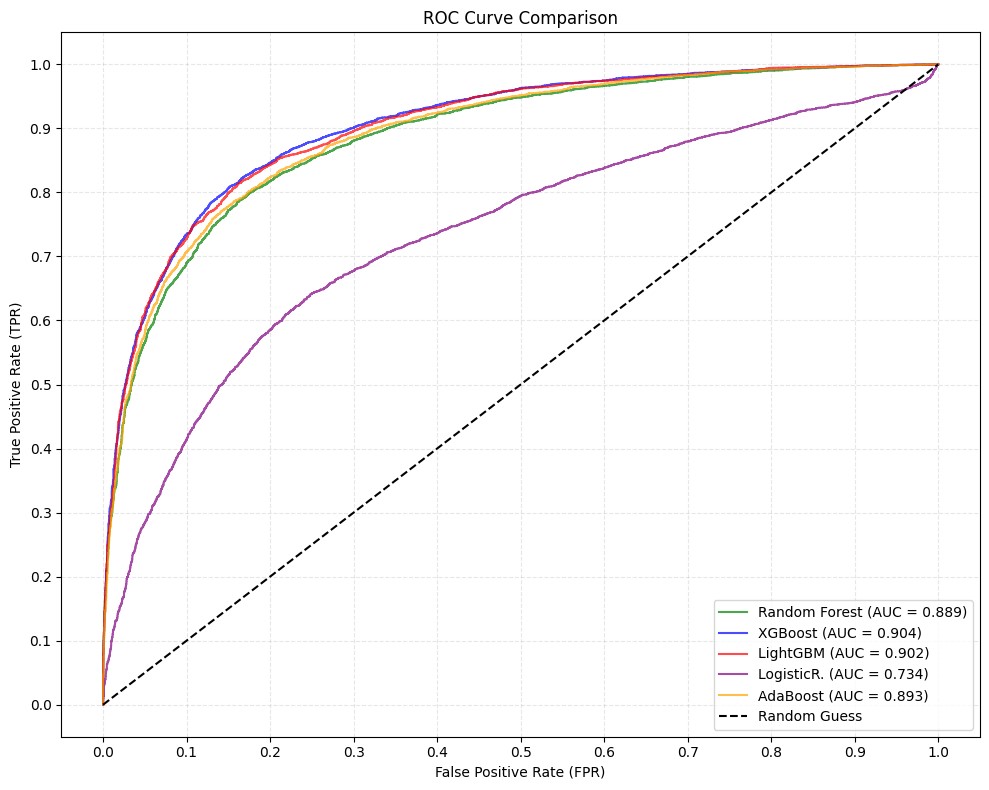}
        \caption{All categories with aerodynamic features}
        \label{fig:roc2}
    \end{subfigure}
    
    \vspace{0.5cm}
    
    \begin{subfigure}[b]{0.48\textwidth}
        \includegraphics[width=0.90\textwidth]{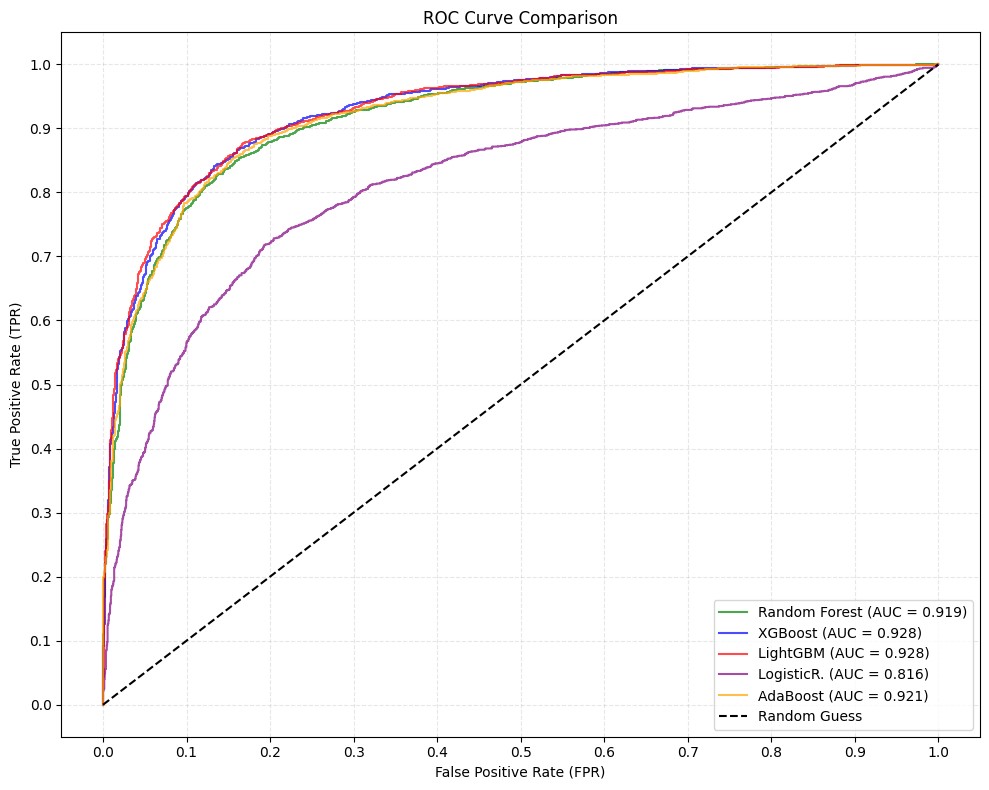}
        \caption{Moderate or greater categories without aerodynamic features}
        \label{fig:roc3}
    \end{subfigure}
    \hfill
    \begin{subfigure}[b]{0.48\textwidth}
        \includegraphics[width=0.90\textwidth]{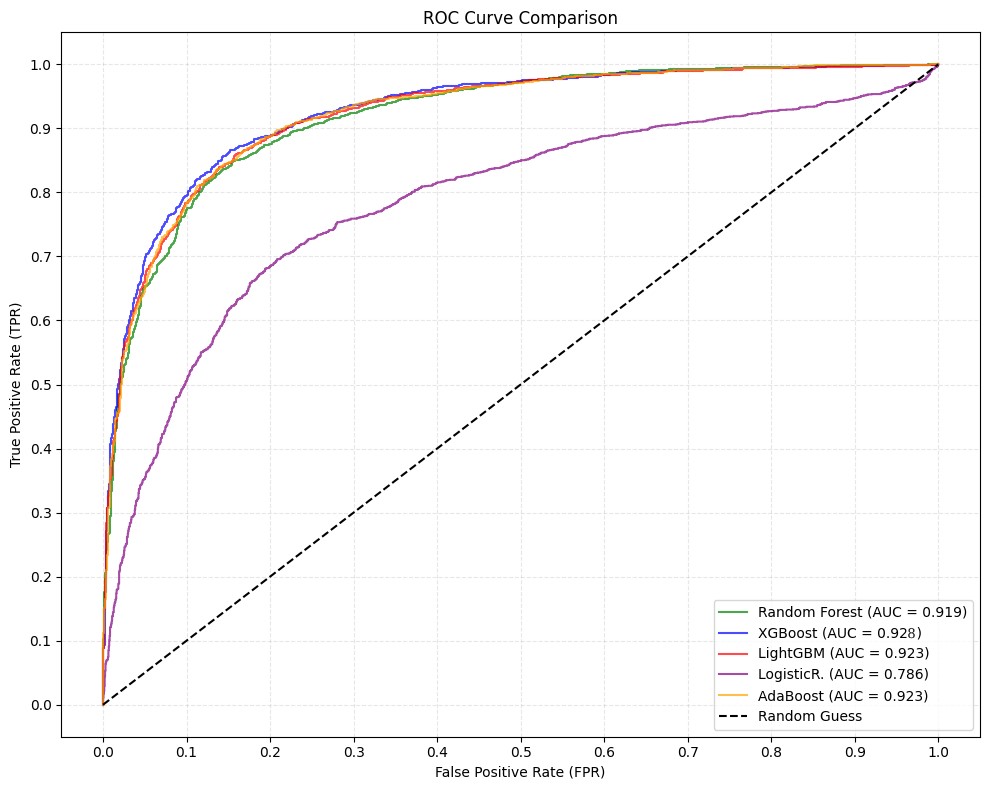}
        \caption{Moderate or greater categories with aerodynamic features}
        \label{fig:roc4}
    \end{subfigure}
    
    \vspace{0.5cm}
    
    \begin{subfigure}[b]{0.48\textwidth}
        \includegraphics[width=0.90\textwidth]{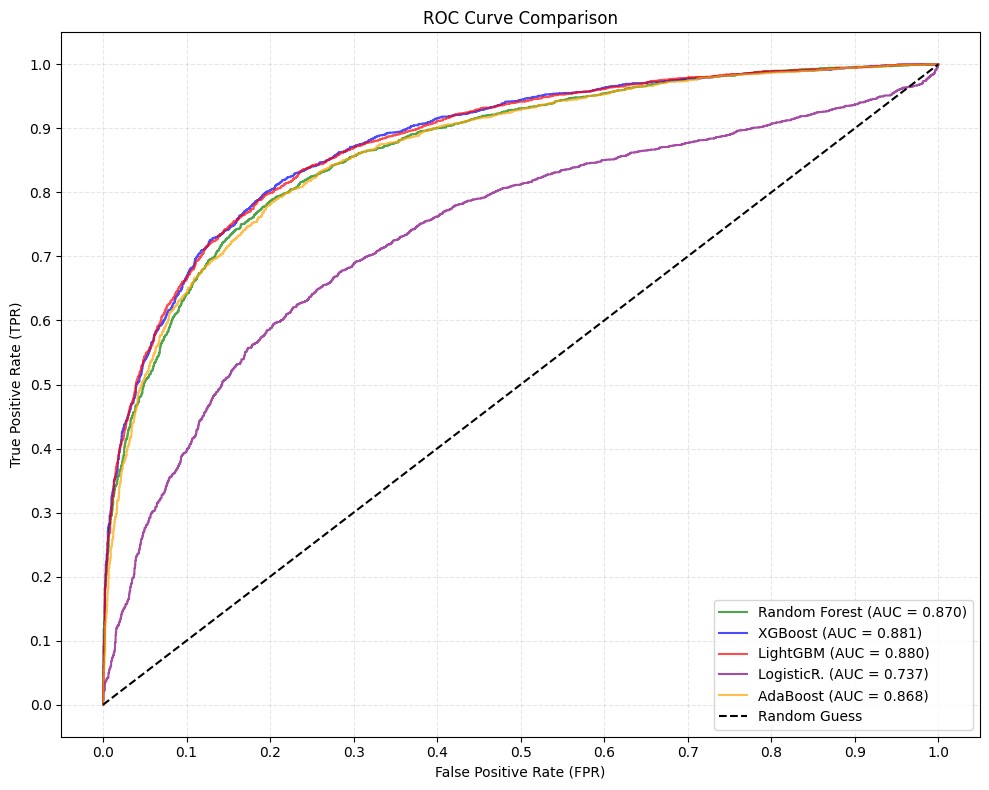}
        \caption{Light and Light MOD categories without aerodynamic features}
        \label{fig:roc5}
    \end{subfigure}
    \hfill
    \begin{subfigure}[b]{0.48\textwidth}
        \includegraphics[width=0.90\textwidth]{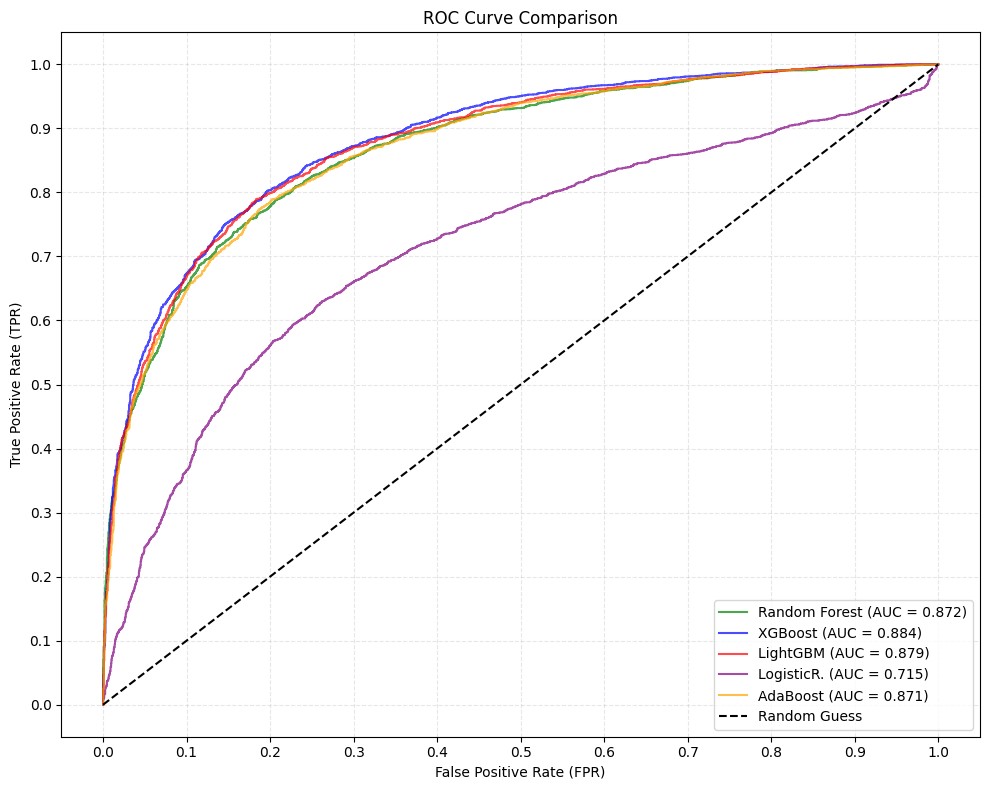}
        \caption{Light and Light MOD categories with aerodynamic features}
        \label{fig:roc6}
    \end{subfigure}
    
    \caption{Comparison of ROC curves on different datasets. Each graph shows the classification performance of different machine learning models (Random Forest, XGBoost, LightGBM, Logistic Regression, and AdaBoost) along with their AUC values.}
    \label{fig:roc_all}
\end{figure}

\end{document}